
\input phyzzx

\overfullrule=0pt
\hsize=6.0in
\vsize=8.9in
\voffset=0.0in
\hoffset=0.0in
\line{\hfill BROWN-HET-892}
\line{\hfill SUTPD/ 93/ 72/ 7}
\line{\hfill April 1993}
\vskip1.5truein
\titlestyle{{GAMMA RAYS FROM SMALL SCALE STRUCTURES ON LONG COSMIC STRINGS
\foot{Work supported in
part by the Department of Energy under
contract DE-FG02-91ER 40688 -Task A}}}
\bigskip
\author{M. Mohazzab\footnote{**}{On leave
from Department of
Physics, Sharif University of Tenchnology, Tehran, Iran} and R. Brandenberger}
\centerline{{\it Department of Physics}}
\centerline{{\it Brown University, Providence, RI 02912, USA}}
\bigskip
\abstract
The formation of cusps on long cosmic strings is discussed and
the probability of cusp formation is estimated. The energy
distribution of the gamma-ray background due to cusp annihilation
on long strings is calculated and compared to observations. Under optimistic
assumptions about the cusp formation rate, we find that strings with a mass per
unit length $\mu$ less than
${G\mu} = 10^{-14}$  will have an observable effect.
However, it is shown that the gamma-ray
bursters can not be attributed to long ordinary strings (or loops).

\endpage

{\bf\chapter{Introduction}}

Grand Unified Theories (GUT) of particle physics predict that the Universe
underwent a series of phase transitions during its early stages of evolution.
In many such theories, the vacuum manifold after symmetry breaking has
nontrivial topology and hence gives rise to the formation of topological
defects after the phase transition. Specifically, in many models the vacuum
manifold has nontrivial first homotopy group, in which case the defects are
linear - cosmic strings [1].

There are various types of cosmic strings. Depending on whether the symmetry
which is broken is a global or local symmetry we get global or local cosmic
strings, and local cosmic strings may be non-superconducting (ordinary) or
superconducting [2]. In this paper we focus on models with ordinary local
cosmic strings. The key parameter in such models is the mass per unit length
$\mu$.
All the purely gravitational consequences of cosmic strings depend only on this
parameter.

For a value of $\mu$ given by
$$G \mu \simeq 10^{-6},\eqno\eq$$
cosmic strings give rise to a successful model for cosmological structure
formation [3]. The above value of $\mu$, determined on the basis of purely
cosmological considerations, coincides with the scale of symmetry breaking in
the simplest GUT models compatible with the running of the coupling constants
of the Standard Model of particle physics [4]. However, in more complicated GUT
models, in particular in models with several stages of symmetry breaking,
cosmic strings may arise with a value of $\mu$ smaller than the one given in
Eq. (1.1).

One of the motivations for this paper is to search for bounds on (and
observational signatures for) models with a value of $\mu$ smaller than the one
given in Eq. (1.1). Another motivation for the research is to check the
consistency of the cosmic string model with the high energy gamma ray
observations (both the high energy background and the frequency of gamma ray
bursters), and to see whether cosmic strings can be responsible for any of the
observations.

The reason why we might expect useful constraints is the following: cosmic
strings typically will lead to cusps [5]. A cusp is a point on the string
which reaches the speed of light and where the derivative of $x^\mu(\sigma, t)$
with respect to $\sigma$ vanishes. Here, $x^\mu(\sigma, t)$ is the string world
sheet, and $\sigma$ is the affine parameter of the string. At a cusp, the two
segments of the string overlap (see below), there is no topological barrier to
decay, and hence we expect most of the energy of the cusp to radiate in a
short, high intensity burst [6] which will include a substantial amount of high
energy electromagnetic radiation. Cusp annihilations will thus give rise to
gamma ray bursts, and will contribute to the gamma ray background.

Previous work has focused on the effects of cosmic string loops. In [7, 8], the
expected contribution of string loops to the neutrino background was
calculated, and in [9], this work was extended to a calculation of the gammay
ray background. In [10], the probability of cusp annihilations on loops giving
rise to observable bursters was estimated and found to be much smaller than the
observed rate of bursters.

In this work, we estimate the effects of cusps on long strings (strings which
have a radius of curvature comparable or larger than the horizon). We analyze
the way in which small scale structure on long strings (in particular kinks)
can give rise to cusps, and we calculate the observational consequences for
both gamma ray background and rate of gamma ray bursters.

Kinks are points on the string where the
tangent vector to the string changes discontinuosly by a non zero angle.
They are usually formed when strings intercommute and cross each
other. Kinks propagate along the string and decay due to
gravitational radiation [11].

One immediate incentive for this work was the recent discovery by the BATSE
instrument on the NASA GRO observatory of a more or less uniform distribution
of
gamma ray bursts in the range of energy $100KeV-1MeV$ [12]. Attributing
the bursts to galactic phenomena like neutron stars colliding
with nomadic  small asteroids is not possible since the distribution
of these objects is not uniform over the sky. However, cosmic strings
have a uniform distribution and it is hence
tempting to speculate whether the gamma ray bursts (GRB) can be attributed to
cosmic string cusp annihilation.

In this paper, we study the effects of cusp annihilation on long strings
on the high energy gamma ray background radiation and on the gamma ray
bursts. We focus exclusively on ordinary (i.e. non-superconducting) cosmic
strings. Bursts from superconducting cosmic strings have been discussed in
[13]. The superconducting cosmic string scenario, however, is now very tightly
constrained by the absence of spectral distortions in the cosmic microwave
background [14].

We propose a new mechanism for cusp formation on long strings: small scale
structures on the long strings like
wiggles or kinks will propagate with the speed of light and collide with each
other. The distribution function for the shape of the small scale structures is
not known. If we assume a random distribution for their
shapes, then the probability of cusp formation is close to $50\%$.

Using the methods developed in [7,9], we calculate the gamma ray background and
the gamma ray burst frequency due to cusp annihilation on long strings. As in
the case of string loops [10] we conclude that our mechanism cannot produce the
observed GRB, the reason being that there are not enough strings sufficiently
close to the observer. However, the mechanism has the potential to enhance the
background gamma ray radiation appreciably.

In Section 2, we review cusps and kinks and discuss
how they are formed.
In the next section, we summarize the analysis of [7,9] of gamma ray emission
by cusp annihilation. In Section 4, we calculate the gamma ray background
produced by the distribution of cosmic strings, assuming that the scaling
solution [15] describes the network of strings.
Section 5 answers the questions concerning gamma ray bursts from long strings.
The final section contains our conclusions.

{\bf\chapter{Cusp Formation on Long Strings}}

 The fields making up the cosmic string evolve according to the equations of
motion derived from the field theory action. For a global $U(1)$ string e.g.,
the action is

$$S= \int d^4x\sqrt{-g}({1\over 2}\partial_\mu \phi
\partial^{\mu}\phi - f(\vert \phi \vert^2- \sigma^2)^2),\eqno\eq$$
\noindent
where $\phi $ is a complex scalar field and f is the self coupling constant.
Assuming that the field configuration corresponds to a string with world sheet
 $\chi^{\mu} (s, \tau)$ in four dimensional space-time, where $s ,\tau$ and
$h_{ij}$ are its world sheet
coordinates and metric, the above field theory action reduces [16] to the Nambu
action for the world sheet:
$$S = \mu \int d^2 s \sqrt{h} \partial_i \chi_{\mu} (s, \tau
) \partial_j \chi^{\mu} (s, \tau ) h^{ij} \eqno\eq$$
where  $i$ and $j$ are world sheet indices. This action to a good accuracy
describes the motion of the string world sheet provided
${w\over R} \ll 1$, where $w$ and $R$ are the width and
curvature radius of the string, respectively.  $\mu$ is the mass
per length of the string. The width $w$ of the string is given in terms of
$\mu$ by $w \sim \mu^{-1/2}$ [17].

In the flat space-time limit, the equation of motion which follows from (2.2)
reduces to

$$\ddot{\chi} (s,\tau)= \chi '' (s , \tau )\eqno\eq$$
\noindent
where $\cdot$ and $'$ are the derivatives with respect to $\tau$ and
$s$, respectively. $\tau$ can be chosen to be time, and the gauge conditions
are
$$\eqalign{ \dot{\chi}\cdot \chi ' & = 0\cr
\dot{\chi}^2 + \chi^{'2} & = 0 }.\eqno\eq$$
\noindent
The general solution of (2.3) can be written as the superposition of a right
and a left moving wave:

$$ \chi^{\mu} (s, \tau ) = {1\over 2} \bigl[ \chi^{\mu}_+ (s
- \tau ) + \chi^{\mu}_- (s + \tau )\bigr]\eqno\eq$$
\noindent
with the gauge conditions yielding

$$\vert \chi_-'  \vert^2 = \vert \chi_+ ' \vert^2 = 1.\eqno\eq$$
\noindent
Here and in the following $\chi$ will denote the spatial three vector
corresponding to the original four vector.

A cusp is a point on the string with $\vert \dot{\chi} \vert = 1$ and
$\chi ' = 0$, i.e. a point where the velocity of the string reaches the speed
of light and where the string doubles back on itself.

As first shown in [5], cusps on cosmic string loops generically arise once per
oscillation time.
The vectors $\pm \chi_{\pm }$ are arbitrary functions of $(\sigma \pm
\tau )$, with the condition (2.6) for their first derivatives.  These vectors
describe  closed curves on the
unit sphere which satisfy the periodicity condition

$$\int \chi_{\pm } ' = 0.\eqno\eq$$
\noindent
As a result, the paths on the
unit sphere will pass through both hemispheres.  Two such curves generically
self intersect which means that a cusp will be formed.

We will now generalize the above analysis to the case of small scale
perturbations on long strings. The small scale structure is a remnant of self
intersections of strings at early times. We expect small scale structure to
survive on all scales larger than a minimum scale determined by the decay of
string excitations via gravitational radiation.

The string solution can still be decomposed into left and right moving waves.
However, the periodicity condition (2.7) is no longer valid. Consider a long
string with a right moving and a left moving perturbation approaching
eachother. In this case, far from the regions of support of the perturbations,
$\chi_{\pm} '$ are at the north and south poles of the unit sphere,
respectively. In the regions of support of the perturbations, $\chi_{\pm} '$
execute closed paths on the unit sphere.

The exact properties of the closed paths on the unit sphere traced out by
$\chi_{\pm} '$ are not known except through numerical simulations,
We will make the assumption that even on small scales the perturbations look
locally like a random walk. This corresponds to a fractal dimension of $2$, a
result not supported by some of the numerical cosmic string evolution
simulations [18], which show a significantly smaller fractal dimension on small
scales. However, it is still unclear if all of the small scale structure on
cosmic strings is treated adequately in the present numerical simulations,
and if nothing else the results of this paper might encourage more detailed
numerical studies of small scale structure on strings.

With our assumption, the distribution of the paths $\chi_{\pm} '$ can be
represented by a function $f(\chi_{\pm})$ which is that of white noise
i.e. $f(\chi'_{\pm}) = f(\theta_{\pm }, \phi_{\pm }) = const.$ or by
normalization considerations

$$f(\theta_{\pm }, \phi_{\pm })={1\over {4\pi}}\eqno\eq$$
\noindent
The cusp condition is

$$\eqalign{\chi_-'\vert_{s=cusp} = -\chi_+'\vert_{s=cusp} & \cr
 {\rm{or}} \  \chi'_- \cdot \chi'_+ + 1= 0}\eqno\eq$$
\noindent
Therefore, the probability that the two independent random paths cross each
other will be

$$P_{cusp} = \int d\Omega_{\chi'_-} \int d\Omega{\chi'_+} f(\chi'_-)
f(\chi'_+) \delta (\chi'_- \cdot \chi'_+ + 1) \eqno\eq$$

It is important to note that the parameter $s$ does not have to be
the
same when the two random walks cross each other, since $\chi_{\pm }$
are traveling waves. Thus,  Eq.(9) is the probability of cusp formation on long
strings.

The integral (9) can be easily done with the result
$$ P_{cusp} = {1\over 2}\eqno\eq$$
\noindent
Therefore, the probability of cusp formation on infinite strings when
two arbitrary wiggles collide is about$ 50{\%}$, provided that our rather
stringent assumptions on the small scale structure are satisfied.

In Figure 1 an example of two travelling waves which form
a cusp is shown. Figure 2 shows the corresponding paths of the two
waves of Fig. 1 on the unit sphere.

{\bf\chapter{Radiation from Cusps}}

Cusps are artifacts of the Nambu approximation for strings. When the curvature
radius of the string becomes comparable to the width (which is the case at a
cusp), the internal structure of the string becomes important. For a cusp of a
cosmic string with finite width $w$, there is a region of comoving length
$\ell_c$ where the two segments of the string overlap (see Figure 3).
No topological criterion prevents this segment of the string from annihilating
into a collection of quanta of the constituent fields of the string. The total
energy of this region of the cusp (and hence [6] the maximal energy which can
be released in the cusp annihilation process) is
$$E_{cusp}\simeq \mu \ell_c \eqno\eq$$
For loops, $\ell_c$ has been calculated in [19]. A field theoretic derivation
of Eq. (3.1) for the
energy radiated by cusp annihilation  is given in  [20].

For long strings, the length $\ell_c$ can be computed in analogy to the case of
loops. In the calculations of [19], the radius $R$ of the loop plays the same
role as the length $l$ of the small scale structures radius. Hence,

$$\ell_c \sim w^{1/3} l^{2/3}.\eqno\eq$$

Cusps annihilate into scalar particles and gauge bosons corresponding to the
fields which are excited in the string. These
particles subsequently decay into jets of lower mass particles, like
quarks, gluons and leptons. The empirical QCD multiplicity function [21]
can be applied  to find the energy spectrum of created particles.
For example, consider the decay of a single cusp into particles with initial
energy $Q_f$. Then, the number distribution of photons
with energy $E=xQ_f (0 \le x< 1)$ produced by the neutral pions is [7]

$${dN\over dE}={15\over 16}{\mu \ell_c \over Q_f^2}\big({16\over
3}-2x^{1\over2}-4x^{-{1\over 2}}+{2\over 3}x^{-{
3\over2}}\big)\big\vert_{x=E/Q_f} \eqno\eq$$
\noindent
Eq.(3.3) will be used later to determine the density of photons from
the  distribution of  cusps.

{\bf\chapter{Gamma Ray and UHE Background}}

As discussed before, cusps on long strings can form when two travelling
waves on the string collide. These travelling waves (called ``wiggles" from now
on) can be produced
by interactions of the string network at earlier times. The number density of
wiggles obeys
the scaling solution [15]. The distribution of wiggles is hence on dimensional
grounds given by

$$K(l ,t)={t\over l^2}\eqno\eq$$
\noindent
where l is the size of the kink and t is the horizon
length. $K(l ,t) dl$ is  the number of small scale wiggles
of length in the interval $[l, l+dl]$ on a long string of length t.

The flux $F(E)$ of the gamma ray background (number density of photons
per unit area and time of energy $E$ per $E$ interval) can be obtained by
integrating in time over the contributions of all the sources (here cusps
annihilations) in the past
in the range of energy which after redshifting corresponds to $[E, E+dE]$:

$$F(E)=\int_{t_{rec}}^{t_0} dt f(t, Ez(t))z(t)^{-3}\eqno\eq $$
\noindent
where $z(t)$ is the redshift factor at time $t$, $E$ the gamma ray energy
and $f(t, zE)$ is the number of photons of energy $Ez$ per unit physical
volume at time $t$ per unit time emitted at $t$. ${t_{rec}}$ is the time
of recombination, when the universe becomes transparent to photons.

 For long strings, the above function $f$  can
be obtained by integrating over the contributions from all small scale
structures with size $l$ smaller than $t$ which give rise to cusp formation
with probability $P_c$.

 $$f(t,zE)={z\over Q_f}{dN\over dx}\vert_{x={zE\over Q}}\int_{l_{min}}^t dl
        K(l,t){1\over l}n_{ls}(t) P_c,
\eqno\eq$$
where ${dN\over dE}$ is the number of photons with energy $E$ produced
by each cusp annihilation, ${1\over l}$ is the frequency that two
wiggles meet each other, $n_{ls} $ is the number of long strings per 3-
volume at time $t$ i.e.
$$n_{ls}={\nu \over t^3},\eqno\eq$$
where $\nu $ is a constant ($\nu \simeq 100 $), $P_c$ is the cusp formation
probability and the lower limit for the integration is
$$l_{min}\simeq \gamma G\mu t,\eqno\eq$$
since kinks with $l< l_{min}$ will be smoothed by gravitational radiation.
 In Eq. (4.5) $\gamma \simeq 100$.

  Using (3.2) and the formula for ${dN\over dE}$ in the limit of $E\ll Q_f$ we
obtain

    $$f(t,zE)={45\over 32}{z\over Q_f^2}({Q_f\over {zE}})^{3/2} {\mu
w^{1/3} \over t^{3+1/3}}{1\over (\gamma G \mu)^{4/3}}.\eqno\eq $$
\noindent
Plugging (4.6) into (4.2) and using the redshift formula
$z(t)=({t_0\over t})^{2/3}$, where $t_0$ is the present time, the
gamma ray energy density flux becomes

$$E^3F(E)={45\over 16}[G ^{-5/6}]
{\gamma^{-4/3}\nu \over {t_0}^{2+1/3}}ln(z_{rec})({Q_f\over
{10^{15}{GeV}}})^{-1/2}({E\over {.1 GeV}})^{3/2} (G\mu )^{-1/2}
(10^{-9}GeV).\eqno\eq$$
For $G=.7\times 10^{-38}GeV^{-2}$, $t_0= 4.73\times 10^{17}sec$( $7.188\times
10^{41} GeV^{-1}$) and
$z_{rec} = 1380$ we will have

$$Log_{10} E^3F(E)= -.63 -{1\over 2}Log_{10}(G\mu)+
{3\over 2}Log_{10}({E\over {.1GeV}})- {1\over 2}Log_{10}({Q_f\over
10^{15}GeV})\eqno\eq$$
\noindent
where the units for $E^3F(E)$ are chosen to be $(eV)^{2}m^{-2}sec^{-1}$.

In Figure 4 we have drawn the curves for Eq. (4.8) for various
values of $G\mu $. It is obvious that in order that the radiation from string
does not exceed the observed background radiation we should have the lower
limit
  $${G\mu} > 10^{-14}.\eqno\eq$$

{\bf\chapter{Gamma Ray Bursts From Long Cosmic Strings}}
  Here, we answer the question whether we can attribute the recently observed
gamma ray bursters to cusp annihilations on
long ordinary cosmic strings.

  It has been shown [9,10] that in order for the cusp annihilations  from
cosmic strings loops to be detected as gamma ray bursts, they should occur very
close to the
detector on cosmological distance scales. The basic reason is that the cascade
of the particle
decay due to cusps annihilation results in a wide angular distribution jet [9].
Therefore, gamma rays from the cusp
annihilation will be distributed in a
wide angle and will be too diluted to be detected unless the distance to the
burst is small. The maximum distance of the burst corresponds to a minimal
emission time $t_{min} = t_0 (1 - \epsilon)$ with $\epsilon \sim 10^{-1}$. The
time difference between $t_0$ and $t_{min}$ will be denoted by $\Delta t$.

For long strings, we can apply the same line of reasoning as for
loops[10] and find the number of detected bursters per unit time to be

$$n_{burst}\simeq (\Delta t)^2 N_k(l,t)\omega_c d_c^2(t_{min}) P_c .\eqno\eq$$
\noindent
Here, $N_k(l,t)$ is the number density of the wiggles with size $l$ at time $t$

$$N_k(l,t) \sim  K(l, t) t^{-3}, \eqno\eq$$
\noindent
$\omega_c $ is the frequency of cusp formation (here
$\omega_c\simeq {1\over l}$) and $d_c(t)^2$ is the area of the
past light cone at time $t$ in comoving coordinates.
Hence,
the number of bursts per unit time will be

$$n_{bursts} \sim (\Delta t)^2 K(t_0, t_0) t_0^{-3}{1\over t_0}
d_c^2(t_{min})P_c. \eqno\eq$$

 Evaluating $d_c$ for radiation dominated universe
$d_c(t)=2t_0^{2/3}(t_0^{1/3}-t^{1/3})$ the above inequality becomes

$$n_{bursts} \sim \epsilon^4 t_0^{-1}P_c .\eqno\eq$$
\noindent
This is comparable to the value for loops.
Therefore, cusp annihilations on long ordinary cosmic strings are not
responsible for GRBs.

{\bf {\chapter {Conclusions}}}

We have studied a mechanism for cusp production on long strings which is based
on small scale wiggles on these strings travelling in opposite directions
meeting eachother and forming a cusp in a similar way to how cusps form on
cosmic string loops. Assuming that the small scale structure on the long
strings has the form of a random walk, we estimate the cusp formation
probability
to be close to
$50 {\%}$. The small scale structure on long strings is due to long string
intersections and loop production in the past. These processes generically
produce traveling waves along the string.

Using an upper limit for the energy produced by cusp
annihilation into mesons, we have calculated the contribution of long
ordinary cosmic string cusp annihilation to the gamma ray background. We find
that the gamma ray background increases as $\mu$ decreases and that we can
therefore set a lower bound on $\mu$. Given our optimistic assumptions about
both energy release from cusps and on the small scale structure, we find this
lower bound to be: ${G\mu} > 10^{-14}$. In particular, we find that cusp
annihilation on long strings may dominates over cusp annihilation on loops.
This is a consequence of the fact that most of the energy density in the string
distribution is in long strings.

We have also estimated the number of gamma ray bursters which can be attributed
 to cusp annihilations on long strings. As is the case of loops [10], this
number is much too small to account for the recent observations.

\

\

{\bf ACKNOWLEDGMENTS}

 We wish to thank Miao Li, Richhild Moessner, Leandros Perivolaropoulos, Andrew
Sornborger and Mark Trodden for useful discussions. One of the authors (M.M) is
grateful to Brown University for hospitality and to the Ministry of Culture and
Higher Education of Iran for financial support.

\

\endpage

\noindent{{\bf References}}

\pointbegin
T. W. B. Kibble, {\it J. Phys.} {\bf A9}, 1387 (1976);
A. Vilenkin and E. P. S. Shellard, "Topological Defects and
Cosmology"(Cambridge Univ. Press, Cambridge, 1993).
\point
E. Witten, {\it Nucl.Phys.}{\bf B 242}(1985) 557.
\point
L. Perivolaropoulos, R. H. Brandenberger and A. Stebbins, {\it Phys. Rev.} {\bf
D41} (1990) 1764.
\point
G. Altarelli, "Precision Electroweak Data and Constraints on New Physics",
(CERN), CERN-TH-6525-92.
\point
T. W. B. Kibble and N. Turok, {\it Phys. Lett.} {\bf B116}, 141 (1982).
\point
R. H. Brandenberger, {\it Nucl. Phys.}{\bf B 293}(1987) 812.
\point
J. H. MacGibbon and R. H. Brandenberger, {\it Nucl. Phys.} {\bf B 33}
(1990) 153
\point
P. Bhattacharjee, {\it Phys. Rev.} {\bf D40}, 3968 (1989).
\point
J. H. MacGibbon and R. H. Brandenberger, {\it Phys. Rev. } {\bf D47} 2283
(1993).
\point
R. H. Brandenberger, A. T. Sornborger and M.Trodden, BROWN-HET-896, {\it Phys.
Rev.} {\bf D}, in press (1993).
\point
B. Allen and R. R. Caldwell, {\it Phys. Rev.} {\bf D 43}, 3173
(1991);{\it Phys. Rev.} {\bf D 43}, R2457 (1991).
\point
S. Howard {(NASA, Huntsville)}, {" Results on gamma-ray bursters from
BATSE"}, lecture, Brown University, 10/29/92; I. Wasserman,
{\it AP. J.}, {\bf 394}, 565-573 (1992).
\point
A. Babul, B. Paczynski and D. Spergel, {\it AP. J.}, {\bf 316}, L49-L54,
(1987); B. Paczynski, {\it AP. J.}{\bf 335}, 525 (1988).
\point
J. Mather, {\it AP. J (Lett.)}{\bf 354}, L37, (1990).
\point
A. Vilenkin, {\it Phys. Rep.} {\bf 121}, 263 (1985).
\point
D. Foerster, {\it Nucl. Phys.} {\bf B81}, 84 (1974).
\point
H. Nielsen and P. Olsesen, {\it Nucl. Phys.}{\bf B61}, 45,(1973).
\point
B. Allen and E. P. S. Shellard, {\it Phys. Rev.} {\bf D 45} (1992) 1898.
\point
D. N. Spergel, T. Piran and J. Goodman, {\it Nucl. Phys.} {\bf B 291}
(1987) 847.
\point
M. Mohazzab, in preparation (1993).
\point
C. Hill, D. Schramm and T. Walker, {\it Phys. Rev.}  {\bf D36}, (1987) 1007.

\

\

\

\

{\bf {FIGURES}}

FIGURE 1. Two incoming waves on a long string in z direction that can
form a cusp when they collide.

FIGURE 2. The unit sphere with paths made by the tips of the vectors
$\pm \chi_{\pm}' $. The two curves intersect and therefore cusps will be
formed.

FIGURE 3. The structure of cusps on cosmic strings. $\ell_c$ is the cusp length
and $w$ is the thickness of the cosmic string.

FIGURE 4. The predicted photon background fluxes by long strings as a function
of
$E$ for various values of $G\mu$, with $Q_f =10^{15} GeV$. The  UHE data points
and the dashed line are  the limits for the gamma-ray background.

FIGURE 5. The predicted photon background fluxes by long strings as a function
of $E$ for various values of $Q_f$, with $G\mu= 10^{-6}$

FIGURE 6. The predicted photon background fluxes by long strings with $Q_f=
\mu^{1/2}$, for various values of $G\mu $.

\end